\documentclass[prl,showpacs,twocolumn,floatfix,letterpaper]{revtex4}

\usepackage{graphicx}

\begin{document}

\preprint{submitted to Phys. Rev. Lett.}

\title{Microscopic analysis of the coherent optical generation 
and the decay of charge and spin currents
in semiconductor heterostructures}

\author{Huynh Thanh Duc,
T. Meier, and S. W. Koch}

\affiliation{Department of Physics and Material Sciences Center,
Philipps University, Renthof 5, D-35032 Marburg, Germany}

\date{\today}

\begin{abstract}
The coherent optical injection and temporal decay of 
spin and charge currents in
semiconductor heterostructures is described microscopically,
including excitonic effects,
carrier LO-phonon and carrier-carrier scattering, as well as
nonperturbative light-field-induced intraband and interband
excitations. A nonmonotonous dependence of the
currents on the intensities of the laser beams is predicted.
Enhanced damping of the spin current relative
to the charge current is obtained as a consequence of spin-dependent
Coulomb scattering. 
\end{abstract}

\pacs{72.25.Fe, 42.65.-k, 72.25.Rb}

\maketitle

One of the
fundamental principles of quantum mechanics
is that superpositions of wave functions
lead to interference phenomena
which depend on the relative phase differences.
In the coherent regime, e.g., shortly
after an external perturbation has generated a nonequilibrium situation,
such quantum mechanical interference effects can be employed
for the coherent control of dynamical processes in atomic, 
molecular, biological, and semiconductor systems 
\cite{r1,r2,r3,r4,bio1,r5}.
Many of the measurements and proposals in this area
make use of the coherent evolution of electronic
excitations induced by specially designed
optical laser pulses, e.g., sequences of phase-locked
or suitably chirped beams.

In semiconductors, the ultrafast 
coherent generation of photocurrents 
using
two light beams with frequencies $\omega$ and
$2\omega$ has attracted considerable attention
\cite{dupont,atan,hache}.
As shown in Ref.~\onlinecite{bhat}, the same
type of interference scheme can also be employed
to create pure spin currents
which are not accompanied by any charge current.
Such spin currents generated on ultrafast time scales
have been observed in semiconductors \cite{stevens1,hubner}
and could be useful for future applications
in the area of spintronics \cite{spintro,rmp}.
As shown recently, 
it is even possible to control photocurrents via
the carrier-envelope phase \cite{ce,ce_josab} which
makes this scheme also interesting for optical metrology.
Furthermore, for disordered semiconductors 
it has been predicted that
sequences of temporally delayed excitation pulses
can be used to induce current echoes \cite{schli}.

The coherent generation of photocurrents in semiconductors
by two light fields with frequencies $\omega$ and
$2\omega$ satisfying
$2 \hbar \omega > E_{\mathrm{gap}} > \hbar \omega$,
where $E_{\mathrm{gap}}$ is the band gap energy,
has first been described in terms of
nonlinear optical susceptibilities which have been 
obtained on the basis of band structure 
calculations \cite{atan,najmaie}.
In this framework, 
the optically-induced intra- and interband
transitions are treated within Fermi's golden rule.
The interference between intra- and interband excitations
leads to electron and hole distributions which are not symmetric
in k-space corresponding to a nonequilibrium situation with 
a finite current.
The dynamics of the generation process has been
analyzed using Bloch equations.
This approach has been applied to 
disordered semiconductors within a
two-band model \cite{schli}
and to ordered quantum wells within a
multiband formalism \cite{martiprb}.
Although the relaxation of the photocurrent by
carrier LO-phonon scattering bulk GaAs has been analyzed
\cite{kral}, in most of the existing publications the
temporal decay of the charge and spin 
currents is still modeled by 
phenomenological decay times \cite{stevens1,hubner,ce,ce_josab}.
The Coulomb interaction among the photoexcited carriers 
has so far not been the focus of particular attention.
However, excitonic effects in quantum wires
have recently been addressed on the Hartree-Fock
level \cite{martidoc}.

In this letter, we present and analyze a microscopic many-body
theory that is capable of describing the dynamical generation,
the coherent evolution, and the decay of charge and spin currents.
Our approach is based on the semiconductor Bloch equations (SBE),
i.e., the equations of motion for the optical polarization
and the carrier populations \cite{bb,schaeferbook}.
The equations nonperturbatively include
the light-field-induced intraband and interband 
excitations without a rotating wave approximation. 
The approach is thus capable of describing the nontrivial dependence of
the initially generated currents on the intensities of the 
incident fields beyond the perturbative regime.
The coherent part of the SBE
contains the Coulomb interaction among the photoexcited carriers
on the Hartree-Fock level, i.e., excitonic effects and Coulombic 
nonlinearities due to energy and field renormalizations. 
As correlation contributions we include 
carrier-LO-phonon and carrier-carrier scattering
at the second Born-Markov level \cite{bb,schaeferbook}.

The dynamical optoelectronic response is analyzed using the
Heisenberg equations of motion for the carrier populations
$n^{e}_{\sigma\mathbf{k}}=\langle
a^\dag_{c\sigma\mathbf{k}}a^{}_{c\sigma\mathbf{k}}\rangle$ and
$n^{h}_{\sigma\mathbf{k}}=1-\langle
a^\dag_{v\sigma\mathbf{k}}a^{}_{v\sigma\mathbf{k}}\rangle$ and the
interband polarization $p^{}_{\sigma\mathbf{k}}=\langle
a^\dag_{v\sigma\mathbf{k}}a^{}_{c\sigma\mathbf{k}}\rangle$. 
Here, $a^\dag_{\lambda\sigma\mathbf{k}}$ ($a_{\lambda\sigma\mathbf{k}}$) creates (destroys)
an electron with wave vector $\mathbf{k}$ and spin $\sigma$ in band $\lambda$.
The resulting SBE including intra- and interband excitations read
\cite{meier94prl,meier95prl}
\begin{eqnarray}
& &(\frac{d}{dt} + \frac{e}{\hbar}\mathbf{E}(t)  \cdot \nabla_\mathbf{k} )\,
p^{}_{\sigma\mathbf{k}} = \left.\frac{d}{dt}p^{}_{\sigma\mathbf{k}}\right|_{\mathrm{coll}}
\nonumber \\
& & -\frac{i}{\hbar} [
\epsilon_{c\mathbf{k}} -\epsilon_{v\mathbf{k}}
- \sum_{\mathbf{q}\neq 0} V_\mathbf{q} ( n^e_{\sigma\mathbf{k+q}}
+ n^h_{\sigma\mathbf{k+q}} )]
 \, p^{}_{\sigma\mathbf{k}} \nonumber \\
& & -\frac{i}{\hbar}(n^e_\mathbf{k}+n^h_\mathbf{k}-1) ( \mathbf{d}_{\sigma\mathbf{k}}^{cv} 
\cdot \mathbf{E}(t)
+\sum_{\mathbf{q}\neq
0}V_\mathbf{q}p^{}_{\sigma\mathbf{k+q}} )  
, \label{sbep} \\
&&( \frac{d}{dt} + \frac{e}{\hbar}\mathbf{E}(t)  \cdot \nabla_\mathbf{k} ) \,
n^\alpha_{\sigma\mathbf{k}} =   \left.\frac{d}{dt}n^\alpha_{\sigma\mathbf{k}}\right|_{\mathrm{coll}}
\nonumber \\
& & -\frac{2}{\hbar}\
{\mathrm{Im}} [ ( \mathbf{d}_{\sigma\mathbf{k}}^{cv} \cdot \mathbf{E}(t)
+\sum_{\mathbf{q}\neq
0}V_\mathbf{q}p^{}_{\sigma\mathbf{k+q}} )
p^{*}_{\sigma\mathbf{k}} ] 
,\label{sbene} 
\end{eqnarray}
where $\alpha=e,h$, 
$V_\mathbf{q}$ denotes the Coulomb interaction potential, 
and $\mathbf{d}_{\sigma\mathbf{k}}^{cv}$ is the interband transition dipole.
The terms given explicitely in Eqs.~(\ref{sbep})-(\ref{sbene}) describe the
dynamics on the time-dependent Hartree-Fock level.
The incoherent contributions
are denoted by $\left. \right|_{\mathrm{coll}}$. 
Here, these {\it collision} terms describe
carrier LO-phonon and carrier-carrier scattering
in the second-order Born-Markov approximation \cite{bb,schaeferbook}.

Except for the explicit inclusion of the spin, equations similar
to Eqs.~(\ref{sbep})-(\ref{sbene}) have been used
to describe the
optoelectronic response of semiconductor superlattices in the
presence of static and Terahertz fields \cite{meier94prl,meier95prl}.
In these studies, the optical fields have been considered
to generate interband transitions via
$\mathbf{E}(t) \cdot \mathbf{d}_{\sigma\mathbf{k}}^{cv}$
and the static and Terahertz fields lead to the intraband acceleration
via $\mathbf{E}(t) \cdot \nabla_\mathbf{k}$. This distinction is
useful if the involved fields
are characterized by very different frequencies.
However, since we here describe the
generation of currents by the interference of optical
fields with frequencies $\omega$ and $2\omega$ 
this simplification
is not possible and thus the total field is considered for
both the intra- and interband excitations.

Numerical solutions of Eqs.~(\ref{sbep})-(\ref{sbene})
provide the time-dependent polarization and
populations.
The populations in the valence and conduction bands
determine the charge and spin current densities which are given by
$\mathbf{J}=e\sum_{\sigma\mathbf{k}}\mathbf{v}^c
n^e_{\sigma\mathbf{k}} - e\sum_{\sigma\mathbf{k}}\mathbf{v}^v
n^h_{\sigma\mathbf{k}}$ and
$\mathbf{S}=\frac{\hbar}{2}\sum_{\sigma\mathbf{k}}\sigma\mathbf{v}^c
n^e_{\sigma\mathbf{k}} -
\frac{\hbar}{2}\sum_{\sigma\mathbf{k}}\sigma\mathbf{v}^v
n^h_{\sigma\mathbf{k}}$, respectively,
where $\mathbf{v}^\lambda = \nabla_\mathbf{k}
\epsilon_{\lambda\mathbf{k}}/ \hbar$ is the group velocity. 

The coherent generation of currents is due to
material excitations which are not symmetric in k-space.
In such situations,
solutions of Eqs.~(\ref{sbep})-(\ref{sbene})
including the scattering contributions are quite demanding.
In order to keep the numerical requirements within reasonable limits
we consider for the present analysis two model systems:
a one-dimensional (1D) quantum wire and 
a two-dimensional (2D) quantum well representative of a GaAs/AlGaAs
system.
The electronic band structure is described in effective mass approximation
using $m_c$=0.067~$m_0$ and $m_v=-$0.457~$m_0$, 
and the band gap is $E_{\mathrm{gap}}$=1.5~eV.
The interband transition dipoles are taken as
$\mathbf{d}_{\uparrow \mathbf{k}}^{cv}=d_{cv}(1,i,0)$ and
$\mathbf{d}_{\downarrow \mathbf{k}}^{cv}=d_{cv}(1,-i,0)$
with $d_{cv}=3~e\AA$, i.e., 
we use
the usual circularly polarized dipole matrix elements which
describe heavy hole to conduction band transitions in quantum
wells close to the $\Gamma$-point \cite{note1}.
The incident laser field is given by
\begin{equation}
\mathbf{E}(t)= \sum_{\nu=\omega,2\omega}
\mathbf{e}_{\nu} A_{\nu}
( e^{-(t/\tau_L)^2}
e^{-i \nu t- i \phi_{\nu}} + c.c.) ,
\label{fields}
\end{equation}
where $\mathbf{e}_{\nu}$ denotes the polarization,
$A_{\nu}$ the amplitude, and $\phi_{\nu}$ the phase of the
field of frequency $\nu=\omega$ and $2\omega$, respectively.
Both frequency components are Gaussian shaped
pulses with a duration
determined by $\tau_L$.
For the case that both
field components are linearly polarized
in x-direction, i.e.,
$\mathbf{e}_{\omega}=\mathbf{e}_{2\omega}=\mathbf{e}_{x}$
the photoexcitation produces a pure charge current since the two
spin systems are excited identically.
For the case of linear perpendicularly polarized pulses,
i.e.,
$\mathbf{e}_{\omega}=\mathbf{e}_{x}$
and $\mathbf{e}_{2\omega}=\mathbf{e}_{y}$,
a pure spin current with no accompanying charge current 
is generated.
Since the charge (spin) current is proportional to
$\sin(\phi^{}_{2\omega}-2\phi^{}_{\omega})$ \cite{hache}
($\cos(\phi^{}_{2\omega}-2\phi^{}_{\omega})$) \cite{bhat}
we use $\phi^{}_{2\omega}-2\phi^{}_{\omega}=\pi/2$
($\phi^{}_{2\omega}-2\phi^{}_{\omega}=\pi$) in our calculations
to obtain maximal currents.

\begin{figure}[t]
\resizebox{8.5cm}{!}{
\includegraphics{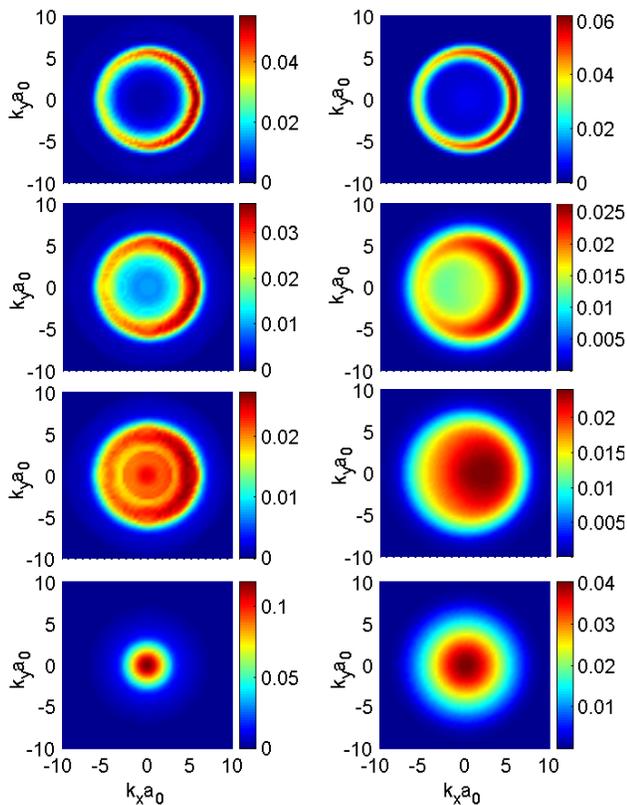}}
\vspace*{-.3cm}\caption{(color) 
Left (right) column: Contour plots of the 
electron (hole) distributions of a quantum well 
in ${\bf k}$-space at $t=$50, 100, 150, and 400 fs (from top to bottom),
respectively.
The incident pulses have a duration of $\tau_L$=20~fs,
the amplitudes are
$A_{\omega} = 2 A_{2\omega}= 108 A_0$, with
$A_0=E_0/ea_0\approx 4\times10^3$~V/cm, where
$E_0$ is the three-dimensional exciton Rydberg,
and $2\hbar\omega=1.65$~eV.
The density of the photoinjected carriers is
$N=10^{11}$~cm$^{-2}$ and
the temperature is $T=50$~K.
$a_0$ is the three-dimensional exciton Bohr radius.}
\label{fig_distributions}
\end{figure}

\begin{figure}[t]
\resizebox{8.5cm}{!}{
\includegraphics{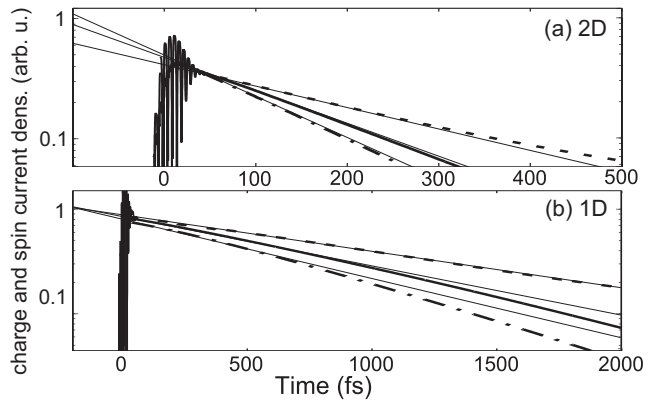}}
\vspace*{-0.3cm}\caption{(a) Time-dependent charge (solid) and spin (dash-dot)
currents of a quantum well 
for the same parameters as in Fig.~\ref{fig_distributions}.
Also shown is the identical decay of both currents if only
carrier LO-phonon scattering is considered (dashed).
The thin solid lines represent exponential decays with time constants of
$240$, $155$, and $125$~fs, respectively.
(b) Same as (a) for a quantum wire.
The density of the photoinjected carriers is
$N=5\times 10^{5}$~cm$^{-1}$ and the other parameters
are the same as in (a).
The thin solid lines represent exponential decays with time constants of
$\tau=1250$, $900$, and $740$~fs, respectively.}
\label{fig_currents}
\end{figure}

Time-dependent electron and hole distributions of a quantum well
in k-space
are shown in Fig.~\ref{fig_distributions}.
The carriers are initially generated with a combined excess
energy of $150$~meV above the band gap. Due their smaller mass
about $130$~meV of the kinetic energy is given to the electrons
and the excess energy of the holes is only about $20$~meV.
Therefore, 
the electron relaxation is considerably influenced by
LO-phonon emission
whereas the hole distribution is only weakly affected.
Immediately after the excitation, the electron and hole
distributions are very similar, see Fig.~\ref{fig_distributions}.
Both are generated nonuniformly on a ring with radius $\approx 5/a_0$.
Due to the quantum interference, the distributions are larger for
positive $k_x$ than for negative $k_x$.
Thus the distributions have a nonvanishing 
positive average momentum which corresponds to a current
in $x$-direction.
In the course of time, the distributions relax towards 
quasi-equilibrium distributions. 
For long times, due to their larger mass the distribution
of the holes is wider than that of the electrons.

Figure~\ref{fig_currents}(a) demonstrates that 
for the considered excitation conditions
both carrier LO-phonon and carrier-carrier scattering contribute significantly 
to the current dynamics. If only carrier LO-phonon scattering is
taken into account, the charge and spin currents decay similarly.
This decay is not exponential, however, its onset can be described by an
exponential decay with time constant $240$~fs.
When carrier-carrier scattering is included, the currents decay
more rapidly and, in particular, the spin current decays faster 
than the charge current.
Additional calculations which omit specific contributions
have revealed that this difference is due to
Coulomb scattering between carriers with different spin.
When a charge current is excited 
($\mathbf{e}_{\omega}=\mathbf{e}_{2\omega}=\mathbf{e}_{x}$),
the electron distributions for different spin are equal
$n^e_{\uparrow \mathbf{k}} = n^e_{\downarrow \mathbf{k}}$.
Thus, in this case both spins have the same nonvanishing
average momentum. Since carrier-carrer scattering only exchanges
momentum among the carriers, this average momentum
is not reduced by the Coulomb scattering. 
The situation is, however, different when a spin current is excited
($\mathbf{e}_{\omega}=\mathbf{e}_{x}$, $\mathbf{e}_{2\omega}=\mathbf{e}_{y}$).
In this case the electron distributions satisfy 
$n^e_{\uparrow \mathbf{k}} = n^e_{\downarrow \bar\mathbf{k}}$
where $\bar\mathbf{k}$ has the same $y$-component as $\mathbf{k}$
but its negative $x$-component, i.e.,
the average momenta of the two spins point into opposite 
directions. Therefore, the total electron momentum vanishes
and Coulomb scattering leads to a decay of the  
average momenta of the spin-up and spin-down electrons.
Fig.~\ref{fig_currents}(b) shows that qualitatively similar results
are obtained for quantum wires.
However, due to the smaller 1D phase space, the scattering is 
reduced and thus the decay times are longer than in 2D.
The results presented in Fig.~\ref{fig_currents} and of additional evaluations
show that 
i) at low densities the charge and spin currents decay with the same time
constant due to carrier LO-phonon scattering and
ii) with increasing density, carrier-carrier scattering 
speeds
up the decay of both currents, however, the rate of change is larger for the
spin than for the charge current. 

\begin{figure}
\resizebox{8.5cm}{!}{
\includegraphics{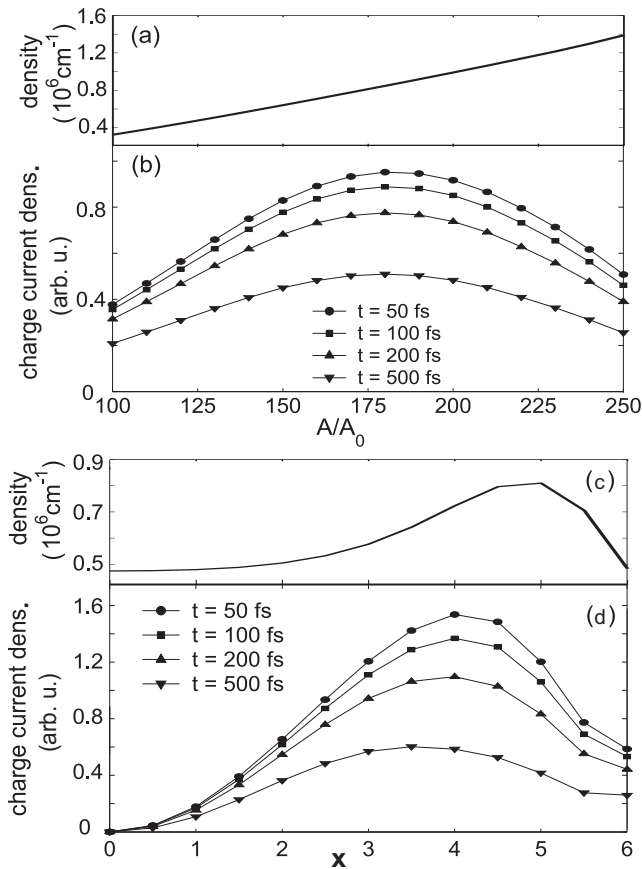}}
\vspace*{-.3cm}
\caption{Dependence of (a) the carrier density and
(b) the charge current density at different times
on the amplitude $A=A_{2\omega}=A_{\omega}/2$
of the incident pulses in a quantum wire.
Dependence of (c) the carrier density and
(d) the charge current density 
on the amplitude ratio
$x=A_{\omega}/A_{2\omega}$ for $A_{2\omega}=128 A_0$.
Here, we use $2\hbar\omega=1.62$~eV and $T=50$~K.}
\label{charge_a0}
\end{figure}

Since 2D calculations are numerically extensive, we present in the
following results obtained in 1D.
Qualitatively similar effects should also be present in 2D.
Fig.~\ref{charge_a0}(a) shows how the injected carrier density
depends on the field amplitude $A=A_{2\omega}=A_{\omega}/2$.
In the considered regime, the density increases approximately
linear with $A$.
However, the current depends on $A$ in a strongly nonlinear fashion,
see Fig.~\ref{charge_a0}(b).
For small amplitudes,
the nonsymmetric k-space carrier distributions
increase with $A$ without significant distortion.
Therefore, in this regime both the total density and
the charge current become larger with increasing $A$.
At a certain excitation level
the peak of the generated distribution 
becomes comparable to $1$ and further
excitation is reduced by the Pauli principle.
Thus, in the high-field regime the 
asymmetry of the k-space distribution and the current decrease
with increasing field amplitude.

Figures~\ref{charge_a0}(c) and (d) show 
how the carrier and the charge
current densities depend on the ratio of the field amplitudes 
$x=A_{\omega}/A_{2\omega}$ for a fixed amplitude
of the $2\omega$ field $A_{2\omega}=128 A_0$. 
At $x=0$, only the
$2\omega$ field is present which generates via interband
excitations symmetric carrier distributions in k-space, i.e.,
no current. 
A finite current is present only
if $x$ is finite since the combined action of inter- and intraband
excitations is required for the generation of nonsymmetric k-space
carrier distributions. 
In the limit of small $x$ the current increases as $x^2$
in agreement with a perturbative analysis of the light-matter coupling.
In this regime,
the carrier density increases only slightly
with $x$ since a weak $\omega$
field predominantly redistributes the carriers in k-space
by introducing asymmetries.
Beyond this perturbative regime we find an interesting
dependence of both the carrier density and the current on $x$.
The largest current is obtained for $x \approx 4$, i.e.,
$A_{\omega} \approx 512 A_0$.
It can be expected that for optical frequencies, 
the $x$ which generates the largest current is
always larger than $1$, since 
intraband excitations are relatively smaller than
interband excitations.
We find that 
the optimal value of $x$ increases with decreasing
$A_{2\omega}$. Thus in the limit of weak fields 
$A_{\omega}$ should be chosen much larger than $A_{2\omega}$
if one is interested in generating large currents.
Our calculations have shown that the intensity dependence 
of the spin current is very similar that of the charge current,
see Fig.~\ref{charge_a0}.

In summary, the coherent optical injection and the decay of 
charge and spin currents in
semiconductor heterostructures is described by a microscopic
many-body theory. We find that
due to Coulomb scattering between carriers of
different spin at elevated excitation
levels the spin current decays more rapidly than the charge 
current. 
An interesting nonmonotonous dependence of the
currents on the intensities of the laser beams is predicted.
These results should stimulate further experimental research
in this direction.

\begin{acknowledgments}
This work is supported by the Deutsche Forschungsgemeinschaft (DFG),
by the Ministry of Education and Research (BMBF),
and by the Center for Optodynamics, Philipps University, Marburg, Germany.
We thank the John von Neumann Institut f\"ur Computing (NIC),
Forschungszentrum J\"ulich, Germany, for grants for extended CPU time
on their supercomputer systems.
\end{acknowledgments}


\begin{thebibliography}{12}
\bibitem{note1}These matrix elements are clearly not realistic for quantum wires.
This model, however, allows us to perform detailed investigations of the 
currents in a 1D system with rather modest numerical requirements. 
The obtained results are useful since the fundamental
dependencies and the differences between the decay of the 
charge and spin currents are similar in 2D.
\end{thebibliography}
\end{document}